# Context sequence theory: a common explanation for multiple types of learning

Mingcan Yu, Junying Wang

**Abstract**

Although principles of neuroscience like reinforcement learning, visual perception and attention have been applied in machine learning models, there is a huge gap between machine learning and mammalian learning. Based on the advances in neuroscience, we propose the "context sequence theory" to give a common explanation for multiple types of learning in mammals and hope that can provide a new insight into the construct of machine learning models.

**Background**

In recent years, machine learning models such as convolutional neural networks(CNN)[1], Long Short-Term Memory(LSTM)[2], generative adversarial network(GAN)[3] and Transformer[4] have been widely used in various fields. Some of these models referred to the principles of brain science including reinforcement learning, visual perception and attention. However, there is still an enormous gap between machine learning models and human learning or mammalian learning. For example, a large number of training samples are not always necessary for mammalian learning in many conditions. And mammalian brain has a strong flexibility and plasticity to support the outstanding transfer learning.

**Multiple types of learning**

Over two thousand years ago, Confucius, the Chinese philosopher, proposed many laws of learning such as "It's important for learning to practice from time to time" and "we can gain new insights through reviewing old material"[5].

In the past hundred years, many scientists provided various new findings and insights into learning. Pavlov trained his dogs and found a type of learning called "classic conditioning", and created many concepts including "acquisition", "extinction", "generalization" ,"discrimination" and "secondary conditioning" to describe properties of learning. Thorndike found his cats showed "trial and error" learning and Skinner detailly investigated the relationship between learning and reinforcement through operant chambers. Tolman tested mice in various mazes and developed the "cognitive map" theory. Köhler observed that a chimpanzee named Sultan had the ability of "insight learning". And the concept of "observational learning" was proposed by Bandura[6].

Based on these studies, a variety of learning types have been observed, described and explained. However, it is still unknown whether all of these types of learning are controlled by a common fundamental mechanism. Here, we try to give a common explanation for multiple types of learning.

**Context sequence theory**

Bartlett's studies on explicit memory showed that past experiences were used in the present as cues that help the brain reconstruct a past event[7]. And Buzsáki proposed that hippocampus, a brain area involved in learning and memory, could represent spacetime as a"sequence generator"[8]. Based on these studies, we propose a theory named "context sequence theory" to explain different types of learning.

The context sequence theory includes several key points as follows:

(1) Every event can be seen as a series of contexts.

(2) Each context can be transformed into a framework with some ambiguous and flexible contents in the brain.

(3) Learning is a process of context reorganization and context sequence formation.

(4) For neutral contexts, the context sequence in the brain is similar to the sequence in the event and the strengths of associations between any two adjacent contexts are nearly equal.

(5) For positive or negative contexts, the associations between them and their previous contexts are enhanced.

(6) Different context sequences can be integrated into one context sequence. By contrast, one context sequence can

be divided into different context sequences.

(7) After learning, when subjects come into the similar event, brain will recall one of contexts and predict next context based on the context sequence.

**Explanation for a variety of learning types**

For classical conditioning

Acquisition is due to the enhancement of the association between the tone context and the meat context and extinction is due to the weakening of this association.

Generalization means that different tones activate the same tone context in the brain and the brain predicts the next meat context. And discrimination means that different tones(tone A and tone B) and the context with or without meat are transformed two context sequences: one is "tone A context-meat context" sequence, the other is only "tone B context".

Secondary conditioning is because "light context-tone context" sequence and "tone context-meat context" sequence are integrated into "light context-meat context". And a recent study supported that it seems to be "light context-meat context" rather than "light context-tone context-meat context" sequence[9].

For operant conditioning

The cats or mice can enhance the association between the "hit action context" and the "reward context" and weaken the association between the "hit action context" and the previous "miss action contexts" in the brain. This is why cats or mice can reach a hit action more quickly after learning in the operant box.

For insight learning

The chimpanzee Sultan has formed the "box context-banana context" sequence and "stick context-banana context" sequence in the brain. And she can combine the box and the stick to reach bananas because these two sequences are integrated into "box context-stick context-banana context" sequence.

For observational learning

The context sequence is also formed in the brain when someone has observed that others did something. The context sequence can represent the event structure and the details of context are flexible. Therefore, when someone comes into the similar situation, the context sequence will be also activated.

For cognitive map

The key point is that how the cognitive map can be formed from context sequences. For example, after free exploration, mice can always choose the shortest path to get food in the following maze based on the cognitive map[6]. In this event, we can use a, b, c, d and e to represent each part context(Fig.1). After exploration, the context sequence "a-b-e", "a-c-e" and "a-d-e" can represent for the start-to-end paths. The "a-b-c(reverse)-c-e" or "a-b-c-d-e" context sequence can be also formed, but the associations between "a" context and "e" context in these context sequences are weaker than those in the above three context sequences. Obviously, "b", "c" or "d" context can be seen as a context sequence consisting of a series of place contexts and the "c" context sequence has the least place contexts, so the association between "a" context and "e" context is the strongest in the "a-c-e" context sequence. This is why mice can choose the shortest path. In a word, various context sequences and their different association strengths build up the cognitive map.

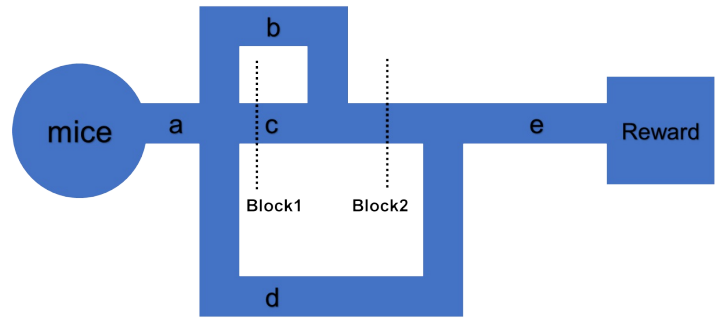

Fig.1 mice transform context sequences to cognitive map in the maze. The diagram is modified from reference 6.

**Support from the recent studies**

Dupret and his colleagues performed single-unit recording in mice in the modified conditioned place preference (CPP) task and showed that high-activity principal cells formed the core of each memory along a first axis, segregating spatial contexts and novelty. Low-activity cells joined co-activity motifs across behavioral events and enabled their crosstalk along two other axes[10]. These findings demonstrate that different contexts and their associations can be formed in the brain.

Another study from Dupret group found that after human and mice had learned the "light-tone" association and the "tone-reward" association, their hippocampus sharp waves represented the "light-reward" association[9]. The evidence support that different context sequences can be integrated into one context sequence, which possibly underlies the secondary conditioning and the insight learning.

Häusser lab found that targeted optogenetic stimulation of specific place cells could trigger remapping of the hippocampal representation of space[11]. This study suggests that contexts in the brain can be reorganized.

**Significance**

Although some of machine learning models such as LSTM involve with sequence data processing, the principles of sequence encoding and processing in the brain have not been used to construct a machine learning model. The "context sequence theory" is not only a common explanation for different types of learning, but also provide a new insight for the design and construction of machine learning models.